\documentclass[sigconf]{acmart}
\usepackage{tabularx,booktabs}

\AtBeginDocument{%
  \providecommand\BibTeX{{%
    \normalfont B\kern-0.5em{\scshape i\kern-0.25em b}\kern-0.8em\TeX}}}

\copyrightyear{2020} 
\acmYear{2020} 
\setcopyright{acmcopyright}\acmConference[MM '20]{Proceedings of the 28th ACM International Conference on Multimedia}{October 12--16, 2020}{Seattle, WA, USA}
\acmBooktitle{Proceedings of the 28th ACM International Conference on Multimedia (MM '20), October 12--16, 2020, Seattle, WA, USA}
\acmPrice{15.00}
\acmDOI{10.1145/3394171.3413536}
\acmISBN{978-1-4503-7988-5/20/10}

\begin{document}
\fancyhead{}

\title{Efficient Adaptation of Neural Network Filter for Video Compression}


\author{Yat-Hong Lam}
\affiliation{%
  \institution{Nokia Technologies}
  \streetaddress{Hatanpaan valtatie 30,}
  \city{Tampere}
  \country{Finland}
  \postcode{33100}
}
\email{yat.lam@nokia.com}
	
\author{Alireza Zare}
\affiliation{%
  \institution{Nokia Technologies}
  \streetaddress{Hatanpaan valtatie 30,}
  \city{Tampere}
  \country{Finland}
  \postcode{33100}
}
\email{alireza.zare@nokia.com}

\author{Francesco Cricri}
\affiliation{%
  \institution{Nokia Technologies}
  \streetaddress{Hatanpaan valtatie 30,}
  \city{Tampere}
  \country{Finland}
  \postcode{33100}
}
\email{francesco.cricri@nokia.com}

\author{Jani Lainema}
\affiliation{%
  \institution{Nokia Technologies}
  \streetaddress{Hatanpaan valtatie 30,}
  \city{Tampere}
  \country{Finland}
  \postcode{33100}
}
\email{jani.lainema@nokia.com}

\author{Miska M. Hannuksela}
\affiliation{%
  \institution{Nokia Technologies}
  \streetaddress{Hatanpaan valtatie 30}
  \city{Tampere}
  \postcode{33100}
  \country{Finland}
}
\email{miska.hannuksela@nokia.com}


\begin{abstract}
We present an efficient finetuning methodology for neural-network
filters which are applied as a postprocessing artifact-removal step
in video coding pipelines. The fine-tuning is performed at encoder
side to adapt the neural network to the specific content that is being
encoded. In order to maximize the PSNR gain and minimize the bitrate
overhead, we propose to finetune only the convolutional layers' biases.
The proposed method achieves convergence much faster than conventional
finetuning approaches, making it suitable for practical applications.
The weight-update can be included into the video bitstream generated
by the existing video codecs. We show that our method achieves up
to 9.7\% average BD-rate gain when compared to the state-of-art Versatile
Video Coding (VVC) standard codec on 7 test sequences.
\end{abstract}




\maketitle

\section{Introduction}

Digital video content is accounted for the majority of media transmission
nowadays. Due to its richness in content, the high volume demand poses
challenges to bandwidth utilization, thus making video compression
an essential technology. However, compression algorithms often generate
artifacts in the decoded video which reduce the visual quality perceived
by observers.

These compression artifacts are similar in many different compressed
video contents so it is possible to suppress them by filtering. Apart
from the traditional filtering approaches, convolutional neural networks
(CNNs) were integrated into the traditional codec to replace the traditional
filters \citep{Jia2019}. CNNs can be also used as a post-processing
filter after the traditional decoding steps \citep{ChaoDong2015,LukasCavigelli2017}.
As for neural network architectures, several works have adapted architectures
which were initially thought for super-resolution tasks, and used
them for reducing compression artifacts \citep{yu2016deep}.

While other works \citep{zhang2018residual,HujunYin2019} utilize multiple
networks to allow some adaptability to the content, we present a novel
approach in this paper to send bias only weight-update as adaptation
signal to achieve the adaptation in a much more efficient way. A basic
post-processing filter is pretrained on a general image dataset to
learn the general types of compression artifacts. The pretrained filter
is incorporated in the decoder side after the traditional decoding
steps. In encoding stage, the filter is adapted on the target video
content by finetuning only the bias terms of the convolutional layers.
The updated coefficients of the bias terms are then encoded and provided
to the decoder.

\begin{figure*}
\hfill{}\includegraphics[width=0.8\paperwidth]{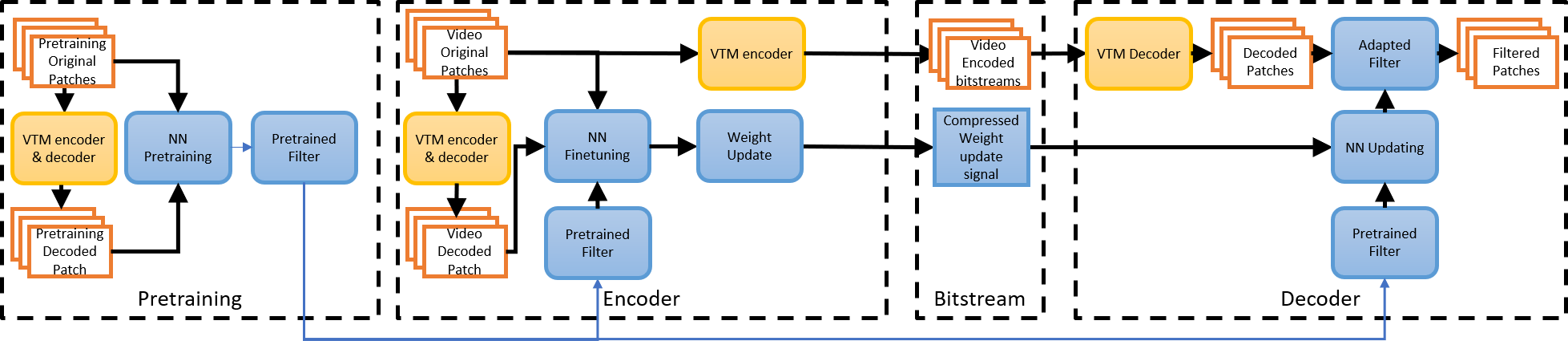}\hfill{}

\caption{\label{fig:Overview}Overview of the encoder-decoder structure. The
traditional path is shown in yellow blocks and our proposed method
is shown in blue blocks.}
\end{figure*}

\section{Related Works}

\subsection{Artifact Removal Filtering in Video Coding}

In the conventional video compression standards, various types of
compression artifacts (e.g., blocking, ringing, contouring effects,
and blurring) appear due to the block-based coding and quantization
structure. The artifact removal filters can be designed either as
out-of-loop/post-processing filtering (i.e., performing at the decoder
end) or in-loop filtering (i.e., performing at both encoding and decoding
loops). In the upcoming video coding standard, Versatile Video Coding
(VVC) \citep{Chen2019}, three in-loop filters namely de-blocking
filter (DBF), sample adaptive offset (SAO), and adaptive loop filter
(ALF), have been designed. The filters are applied sequentially in
a pre-defined order as DBF, SAO and ALF. The recent studies demonstrated
great success in applying CNN-based
methods to outperform the conventional filters in suppressing compression
artifacts. CNN-based methods were proposed to replace the conventional
filters \citep{wang2019attention,Jia2019} or integrated to be used
along with the traditional filtering chain \citep{park2016cnn}.

\subsection{Transfer Learning of Filter}

In ALF in-loop filter \citep{chen2012adaptive}, in order to improve
the adaptive capacity of the filter, the filter coefficients are trained
at the encoder side and transmitted to the decoder. Similarly, in
CNN-based methods finetuning process allows a neural network to adapt to the context
by updating existing weightings with new training data. One
common technique used when finetuning part of a neural network is
layer freezing, which allows to update only part of the weights and
keep other weights unchanged. It is commonly used for example in transfer
learning to retain the knowledge from the original network. In some
popular computer vision tasks, such as classifiers or object detectors \citep{huang2017speed}, the initial layers
of the network are kept unchanged and only the last layers are finetuned
to adapt to the new task. Instead of transferring the knowledge to
a new domain or a new task, we want to finetune the pretrained knowledge
of image enhancement to the target content which is in the same domain
and for the same task. The adaptation does not require a topological
change of the network but finetuning in weighting values. In our research,
we demonstrate that this finefuning can be done more efficiently through
bias-only updating and freezing the other weights of convolution layers.

\subsection{Neural Network Representation of Filter}

In \citep{tung2017fine}, the authors propose a method for jointly
finetuning and compressing a pretrained neural network to adapt the
network to a more specialized domain than the pretraining domain,
in order to avoid overfitting due to over-parameterization. However
they have to send the whole neural network which is too costly in
our application. In \citep{yatcompressing}, the authors suggest
an efficient way to reduce the size of adaptation signal of postprocessing
filter for image compression. The finetuning of postprocessing neural
network is conducted with compressive training. However, that work
focuses on image compression and there is no consideration on bandwidth
limitation which is crucial for video compression and is a much more
time-consuming process, as it considers all the weights and adds one
more training loss term. In our approach, we aim to send adaptation
signal in an even more efficient way, by treating the weights of a
layer and its bias terms differently.

\section{Methodology\label{sec:Methodology}}
The outstanding performance in suppressing compression artifact of CNN-based filters depends on its numerous hidden layers and neurons, which means a huge amount of parameters. The single-network based methods have a limited adaptation capacity and require high bitrate to extend their adaptation power, because of the high number of parameters to update. In this work, an efficient and highly adaptive filtering approach is proposed which is based on computing an adapted network for each adaptation interval. In the encoder, the adapted network is computed by performing a finetuning process using a pretrained network which has a general knowledge about different types of artifacts. Then a light weight update signal is generated and transmitted to the decoder side by comparing the adapted and pre-trained networks. The update signal is limited to bias-only parameters to reduce the signalling overhead, while preserving a great deal of adaptation potential. In the decoder side, the adapted network is reconstructed using the update signal and the exact pretrained network embedded in the decoder.

\subsection{Overview}

The proposed method consists of using a traditional codec empowered by a post-processing neural network filter. The overall pipeline is shown in Fig. \ref{fig:Overview}. The traditional components include encoder and decoder with the in-loop filters enabled, which are shown in yellow blocks. A pretrained neural filter is obtained by offline training on a large dataset. The pretrained filter is included in both encoder and decoder. On encoder side, the pretrained filter is finetuned with the target video sequences. The finetuned network can be represented by the weight-update which is transferred with the encoded bitstream. On the decoder side, the adapted filter is restored by the built-in pretrained filter and weight-update. Finally the adapted network is applied on the decoded frames at the decoder end as an out-of-loop filter.

\subsection{Weight-update as Adaptation Signal}

A traditional way of network adaptation is to clone the pre-trained
network and continue training the network on the specific domain.
However this process involves retraining of large amount of parameters, which
is a computational expensive process. Furthermore, as the amount of
parameters grows with the depth and the number of filters in CNN networks,
this naïve approach is not feasible with large networks, as the weight-update
to be encoded would be prohibitively large. In \citep{yatcompressing},
an updated neural network can be represented as an addition of a pretrained
approximation neural network and an updating signal representing the
adaptation. In such manner, only the difference between the pretrained
network and finetuned network, which is significantly smaller in storage
size, is used for the update. Our approach is inspired by them to
represent the finetuned network with weight-update.

By sacrificing the adaptive power of neural network, the size of weight-update
can be further reduced by imposing constraints on the finetuning process.
The finetuning process of \citep{yatcompressing} is carried out with
a compressive loss. It attempted to reduce the size of adaptation
signal with compressive training that avoids unnecessary change in
neural network. However, the introduced subsequent quantization process brings
extra computational burden. Furthermore, the bit rate of the resulted weight update signal is high for video coding application as the weight-update signal involves a lot of parameters from two-dimensional convolutional layers.

Our breakthrough lies on the bias-only adaptation on top of a well
pretrained neural network. Our hypothesis is that the noise-removal
knowledge is stored in the local structure of two-dimensional convolution
kernel, which should keep unchanged in the finetuning process. Only
bias in the neural network should be updated in the finetuning process.
Even under this restriction, significant visual quality improvement
can be obtained. Compared with \citep{yatcompressing}, it is a straight
forward approach to constrain the finetuning scope. It involves much
less parameters being trained compared to updating the whole neural
network (Table \ref{tab:Distribution-of-parameters}). Consequently, the computational load and signalling overhead are significantly reduced and the neural network converges faster.

\begin{table}
\caption{\label{tab:Distribution-of-parameters}Distribution of parameters
in the neural networks with different number of filters $N_{filters}$
and number of blocks $N_{blocks}$}

\begin{tabular}{|ll|lll|l|}
\hline 
\multicolumn{2}{|l|}{Network Structure} & \multicolumn{3}{l|}{No. of parameters} & \tabularnewline
$N_{filters}$ & $N_{blocks}$ & Conv & Bias & Total & Bias/Total\%\tabularnewline
\hline 
512 & 5 & 7109632 & 2563 & 7112195 & 0.036\tabularnewline
512 & 4 & 4750336 & 2051 & 4752387 & 0.043\tabularnewline
256 & 6 & 2965248 & 1539 & 2966000 & 0.052\tabularnewline
256 & 5 & 2375424 & 1283 & 2376707 & 0.054\tabularnewline
128 & 7 & 746115 & 771 & 745344 & 0.103\tabularnewline
128 & 6 & 598531 & 643 & 598531 & 0.107\tabularnewline
64 & 7 & 188739 & 387 & 188352 & 0.205\tabularnewline
64 & 6 & 151811 & 323 & 151488 & 0.213\tabularnewline
\hline 
\end{tabular}
\end{table}

\subsection{Pretraining Stage}


From the prospective of machine learning, the pretrained neural network
should be a good approximation of the finetuned network in order to minimize the size of adaptation signal. In our application the approximation
is possible due to the similarity of compression artifact of a specific
traditional codec. We desire a pretrained network to have a good
generalized knowledge on the specific codec behaviour.

The pretraining stage aims to train a post-processing filter that is embedded in the traditional codec on both encoder and decoder sides. In our methodology, the pretraining stage is
particularly important as the pretrained two-dimensional
kernel will be used directly at the decode end.  The pre-trained neural network should
be trained on a large and diverse content. Ideally, the training data should cover different variety of content in different intended 
applications. Different compression artifacts generated by a codec can be created by encoding and decoding the training data. The original and decoded version of the data are used pairwise
for network training. The codec setting should be identical to the intended application
so the the compression artifacts are similar to those in application
case. The training optimizes quality of processed
data which is evaluated by a quality index. The implementation of
pretraining, including the traditional codec, dataset and training
parameters will be discussed in \ref{subsec:Implementation-details}.

\subsection{Finetuning Stage and Representation of Weight-update}

In the finetuning stage, we want to specialize our neural network
to improve the visual quality of a target video sequence. According
to our hypothesis, the weights of convolutional kernels learned on
a general image database should also work on the test video set, therefore
it is unnecessary to retrain them. The only changes which are required
in order to adapt the pre-trained model to the target video is the bias terms of the model.

The finetuning process takes place online, after traditional encoding process which matches with the one in pre-training process. In a similar manner as in the pretraining stage, the pairwise original and content are used to finetuning the pre-trained neural network. The only exception is that the weights of the convolutional kernels are frozen except for the bias terms, which are finetuned. The obtained weight-update
is then compressed and included into the bitstream.

At decoder side, the adapted neural network is updated by replacing
the bias term of the embedded pretrained network with the decompressed
weight update signal. The adapted filter is applied on the decoded content for removing the compression artifacts. The bias terms of the pretrained network are updated based on an adaptation interval which is configurable.

\section{Experiments}

To prove our hypothesis in Section \ref{sec:Methodology}, the methodology
is implemented on a video compression task and evaluated on different video sequences
from the Joint Video Exploration Team (JVET) common test conditions
and evaluation procedures \citep{conditions2012software} to improve
the performance of a traditional video codec. The implementation and evaluation details
are discussed in the following subsections.

\subsection{Implementation Details \label{subsec:Implementation-details}}

The methodology is based on the adaptive training of a post-processing
filter. The aim of the filter is to improve the visual quality of
decompressed video sequence, which is evaluated by the difference between
decompressed frames and their corresponding source frames in terms of
peak signal-to-noise ratio (PSNR). However, it is also preferable
to minimize the size of weight-update which is transmitted to the
decoder. Thus, the goal of the task is to decrease the BD rate in Bjontegaard metrics of rate-distortion-curves (RD-curve) \citep{bjontegaard2001calculation} which reflects a better
trade-off between the bitrate and distortion. Particularly, a negative BD-rate
and a positive BD-SNR indicate an improvement of rate-distortion performance
by our approach.

\subsubsection{Traditional codec and video dataset}

Although many existing video codecs are available, the traditional
video codec used in our experiment is the state-of-art Versatile Video
Codec (VVC) H.266 Test Model VTM-7.0. The test model takes an uncompressed
raw video or image as input. The test video clips are in YUV 4:2:0
color sampling at 8 bits per sample, which are coded with the quantization
parameter (QP) values of 22, 27, 32, and 37 as defined in JVET common
test conditions (CTC). In order to allow the neural network to perform well during testing phase, the encoding and decoding settings
should be similar when encoding/decoding the pretraining data and the test data. In our experiments, we used the same settings. 

\begin{figure}
\includegraphics[width=0.7\columnwidth]{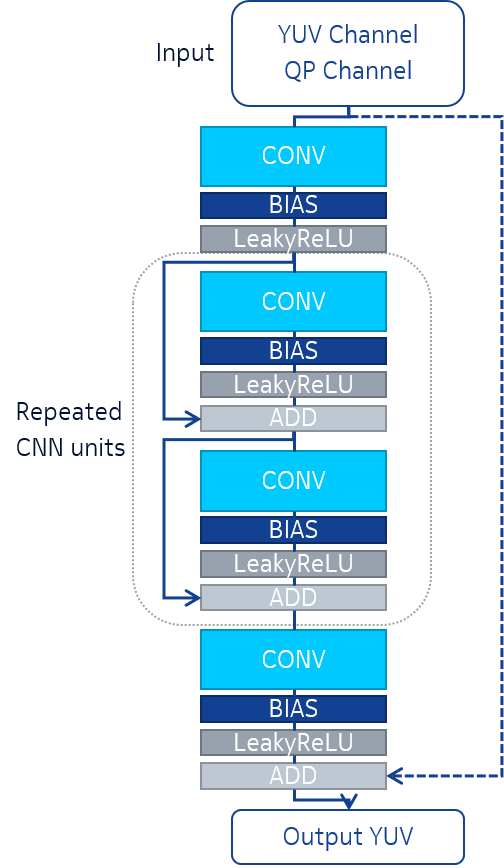}

\caption{\label{fig:The-structure-of}The structure of the post-processing
filter with four convolutional layers. In our implementation, bias terms were separated from the convolutional layer and included in a separate layer. CONV is a convolutional layer excluding the bias terms. BIAS is a layer including only the bias terms of the previous convolutional layer. The number of filters in each
CONV layer and the number of repeated units (2 in the shown example)
are hyperparameters.}
\end{figure}

\subsubsection{Neural Network Structure and Loss Function\label{subsec:Neural-Network-Structure}}

The deep learning part of our approach is implemented in Keras with Tensorflow backend. The network architecture is an auto-encoder which is shown in Fig.
\ref{fig:The-structure-of}. We consider patches (i.e., crops) obtained from the video frames. The input consists of four channels:
the Y channel of decoded patches, the upsampled U and V channels (as input
data is given in the format 4:2:0), and also the corresponding Quantization
Parameter (QP) value used to encode the patches. Higher QP indicates a stronger
compression, thus compression artifacts are more significant. This information is readily available in our application
scenario and providing this extra channel improves the training performance
by learning the knowledge about quantization applied to the content. The QP is normalized and used as the fourth channel of the input.  The output
of the neural network network consists of three channels representing the YUV channels.

The neural network is formed by blocks consisting of a layer that
includes the convolutional kernel's weights (excluding the bias terms),
a layer that includes only the bias terms, and an activation layer 
of type leaky rectified linear unit (LeakyReLU). The number of convolutional
filters, $N_{filters}$, is same for all blocks. The convolutions are applied with stride 1, so the size of the output at each block is same as the size of its input. The kernel size of the convolutional layers is $\left(3,3\right).$
There are skip connections (solid blue arrows) between the input of the blocks
and the output of the blocks. There is another global skip connection (dashed blue arrow)
from the input to the output of the whole network. The output consists of three
channels of same resolution, i.e., the Y, U and V channels. In general,
the trainable parameters in the neural network are those from the
convolutional kernels and from the bias layers. In the pretraining stage,
all the parameters are trained. In the finetuning stage, only the parameters
in bias layers are trained. In our research, there are two main hyperparameters
related to the network architecture: the number of blocks and number
of convolutional filters in the convolutional layer of each block, for which the effect will be discussed.
The characteristics of the considered structures along with the number
of trainable parameters (divided into convolutional kernel weights
and biases) are shown in Table \ref{tab:Distribution-of-parameters}.

The post-processing filter is trained to remove the artifacts and
restore the frame with better visual quality. We aim to optimize the
peak signal-to-noise ratio (PSNR) of the filtered frames, thus the
mean squared error (MSE) between filtered patches and original patches
is used as the training loss. PSNR is used as the evaluation metric, which is commonly used in video compression, for example as one of the
reported metrics in JVET common testing condition. PSNR is also used
in BD-rate calculation which is the metric we desired to optimize
in our task.

\subsubsection{Pretraining Stage}

For the pretraining stage, the 1633 images in Portable Network Graphic
(PNG) format from the training dataset of the Competition on Learned
Image Compression 2020 (CLIC) \citep{CLIC2019} is used as input.
The dataset provides images of various contents and scenes, which
is good for training an initial version of the filter. The images are first
encoded by the VVC/H.266 test model using All Intra (AI) configuration.
The corresponding decoded images and original images are used as training
samples (input-target pairs) of the neural network model.

The neural network post-processing filter is pretrained offline. Patches of size $\left(128,128\right)$ are cropped randomly over
VVC decoded images. These patches are used as training input to
the neural network, whereas the corresponding patches obtained from
the original uncompressed images are used as the ground-truth. During
training, each batch contains 64 patches from 64 different random images in the
dataset. At every training epoch, 32768 patches (in 512 batches) are used. The pretraining process is a long process that takes 2000 epochs,
but the total training time depends on the size of the neural network.
The learning rate is controlled by the Adam optimizer \citep{kingma2014adam}, with
initial learning rate of 0.001.

\subsubsection{Finetuning Stage}

In the finetuning stage, the 7 different video sequences defined in
the Joint Video Exploration Team (JVET) common test conditions and
evaluation procedures \citep{conditions2012software} are used. The
sequences are encoded using Random Access (RA) configuration with
the first 49 frames for each sequence. The frames are broken down
into patches of $\left(128,128\right)$ before the finetuning process.
The patch size of $\left(128,128\right)$ is chosen as the largest
available size in a coding tree unit (CTU) defined in the VVC codec.
The decoding refresh type was set to instantaneous decoding refresh
(IDR) picture with a period of 16. The sequences have resolution
of either 832x480 or 1920x1080, thus belonging to group C of group B
video sequences of JVET dataset, respectively.

The finetuning process is similar to the pretraining process except
that only bias terms are updated. In each epoch, all patches (decoded-original
pair) are used in the finetuning. The learning rate is also controlled
by Adam optimizer but it was decayed to half of its value every 20 epochs until epoch 110. This decay showed to be successful empirically, in terms of better training 
performance and faster convergence.

\subsubsection{Packaging and Decoding Stage}

The resultant bias terms are flattened to a one-dimensional array of 64 bits floating point numbers and
further packaged with 7z, which is a popular lossless compression method based on LZMA2 \citep{LZMA, 7z}. The compressed package is subsequently decompressed in the decoder side. The original bias terms in the pretrained network are replaced with the updated ones from the signaling, while the convolution layer terms remained unchanged. The updated neural network is used to filter the VTM-decoded patches. 

\begin{table*}
\caption{\label{tab:BD-Rate-and-BD-SNR-1}BD-Rate of filtered patches compared
to the VTM decoded patches. A negative BD-rate indicates an improvement
of quality and the best configurations of each dataset are highlighted
in bold letters. The network is labelled as $N_{filters}\_N_{blocks}$ }

\centering
\begin{tabular}{| *{10}{c|}}
\hline 
&  & \multicolumn{4}{c|}{Pretrained BD-rate} & \multicolumn{4}{c|}{Finetuned BD-rate} \tabularnewline
\hline 
Dataset & Network & YUV & Y & U & V & YUV & Y & U & V\tabularnewline
\hline 
B\_BasketBallDrive & 512\_5 & -2.00 & -1.25 & -2.27 & -6.20 & \textbf{-4.35} & \textbf{-2.42} & \textbf{-9.60} & \textbf{-10.72}\tabularnewline
 & 512\_4 & -1.46 & -0.93 & -1.75 & -4.36 & -4.00 & -2.15 & -9.01 & -10.10\tabularnewline
 & 256\_6 & -2.05 & -1.49 & -2.38 & -5.03 & -3.65 & -2.57 & -6.90 & -6.89\tabularnewline
 & 256\_5 & -1.90 & -1.30 & -2.37 & -5.03 & -3.98 & -2.62 & -7.56 & -8.55\tabularnewline
 & 128\_7 & 1.80 & 1.95 & 4.42 & -1.73 & -3.29 & -2.45 & -4.92 & -6.71\tabularnewline
 & 128\_6 & 1.51 & 1.18 & 7.65 & -2.71 & -3.18 & -2.22 & -4.89 & -7.19\tabularnewline
 & 64\_7 & 2.19 & 1.51 & 7.61 & 0.84 & -1.61 & -1.68 & -0.11 & -2.71\tabularnewline
 & 64\_6 & 1.66 & 1.66 & 3.67 & -0.36 & -1.58 & -1.51 & -0.46 & -3.15\tabularnewline
\hline 
B\_BQTerrace & 512\_5 & -3.10 & -2.57 & -1.10 & -8.26 & \textbf{-9.74} & \textbf{-6.76} & \textbf{-17.09} & \textbf{-20.24}\tabularnewline
 & 512\_4 & 1.35 & -0.35 & 10.22 & 2.69 & -9.35 & -5.92 & -17.31 & -21.92\tabularnewline
 & 256\_6 & -2.52 & -2.26 & 3.03 & -9.58 & -7.80 & -6.00 & -9.97 & -16.41\tabularnewline
 & 256\_5 & -3.18 & -2.66 & 0.91 & -10.41 & -8.13 & -6.06 & -12.09 & -16.56\tabularnewline
 & 128\_7 & 10.74 & 11.05 & 17.80 & 1.77 & -5.58 & -5.27 & -4.39 & -8.60\tabularnewline
 & 128\_6 & 12.05 & 10.93 & 19.30 & 11.48 & -5.15 & -4.89 & -4.05 & -7.82\tabularnewline
 & 64\_7 & 12.24 & 9.87 & 18.19 & 20.52 & -3.41 & -3.67 & -1.59 & -3.67\tabularnewline
 & 64\_6 & 8.22 & 8.57 & 16.97 & -2.61 & -3.81 & -3.52 & -2.73 & -6.65\tabularnewline
\hline 
B\_Cactus & 512\_5 & 1.19 & 0.04 & 3.51 & 5.78 & \textbf{-7.28} & \textbf{-3.01} & \textbf{-23.63} & \textbf{-16.53}\tabularnewline
 & 512\_4 & 0.42 & 0.23 & -4.00 & 5.95 & -6.52 & -2.69 & -22.34 & -13.70\tabularnewline
 & 256\_6 & 1.09 & 0.66 & -1.92 & 6.64 & -6.54 & -3.06 & -21.25 & -12.73\tabularnewline
 & 256\_5 & 1.56 & 0.86 & -0.49 & 7.80 & -6.19 & -3.14 & -17.59 & -13.05\tabularnewline
 & 128\_7 & 8.60 & 6.17 & 4.18 & 27.59 & -4.90 & -2.98 & -13.02 & -8.29\tabularnewline
 & 128\_6 & 6.39 & 4.90 & 8.57 & 13.15 & -5.11 & -2.82 & -14.40 & -9.60\tabularnewline
 & 64\_7 & 9.20 & 5.62 & 7.48 & 4.71 & -4.00 & -2.32 & -13.55 & -4.55\tabularnewline
 & 64\_6 & 4.77 & 3.96 & 3.35 & 11.00 & -3.21 & -2.06 & -10.02 & -3.28\tabularnewline
\hline 
\hline 
C\_BasketBallDrill & 512\_5 & \textbf{-2.70} & \textbf{-2.51} & \textbf{-3.64} & \textbf{-2.90} & -1.16 & 1.33 & -8.92 & -8.33\tabularnewline
 & 512\_4 & -2.15 & -2.02 & -1.57 & -3.54 & -1.80 & 0.69 & -8.95 & -9.54\tabularnewline
 & 256\_6 & -2.27 & -2.28 & -2.60 & -1.84 & -1.49 & -0.18 & -5.55 & -5.26\tabularnewline
 & 256\_5 & -2.06 & -2.02 & -1.76 & -2.60 & -2.17 & -0.77 & -7.53 & -5.24\tabularnewline
 & 128\_7 & 4.07 & 4.18 & 4.82 & 2.63 & -2.41 & -1.10 & -6.42 & -6.23\tabularnewline
 & 128\_6 & 3.39 & 2.06 & 7.59 & 7.22 & -1.69 & -1.13 & -4.57 & -2.20\tabularnewline
 & 64\_7 & 3.79 & 1.75 & 11.29 & 8.48 & -0.77 & -1.22 & 0.04 & 1.10\tabularnewline
 & 64\_6 & 3.09 & 1.65 & 7.14 & 7.65 & -1.04 & -0.97 & -2.56 & 0.09\tabularnewline
\hline 
C\_BQMall & 512\_5 & \textbf{-4.29} & \textbf{-3.55} & \textbf{-6.91} & \textbf{-6.17} & -0.38 & 0.99 & -6.08 & -2.88\tabularnewline
 & 512\_4 & -3.69 & -3.06 & -4.37 & -6.80 & -1.06 & 0.38 & -5.61 & -5.08\tabularnewline
 & 256\_6 & -3.63 & -3.37 & -4.62 & -4.19 & -1.07 & -0.54 & -4.25 & -1.03\tabularnewline
 & 256\_5 & -3.80 & -3.32 & -5.07 & -5.44 & -1.91 & -1.09 & -5.85 & -2.83\tabularnewline
 & 128\_7 & 0.41 & 1.04 & -0.88 & -2.11 & -2.03 & -1.66 & -3.90 & -2.34\tabularnewline
 & 128\_6 & 0.22 & 0.21 & 1.66 & -1.15 & -1.78 & -1.62 & -2.74 & -1.82\tabularnewline
 & 64\_7 & 0.95 & 0.17 & 3.53 & 3.04 & -1.30 & -1.66 & -0.66 & 0.18\tabularnewline
 & 64\_6 & 0.63 & 0.28 & 3.51 & -0.19 & -1.14 & -1.50 & -0.31 & 0.22\tabularnewline
\hline 
C\_PartyScene & 512\_5 & -2.37 & -2.48 & -5.96 & 1.91 & -3.97 & -1.95 & -16.16 & -3.89\tabularnewline
 & 512\_4 & -1.99 & -2.02 & -5.97 & 2.19 & \textbf{-3.98} & \textbf{-1.82} & \textbf{-17.19} & \textbf{-3.76}\tabularnewline
 & 256\_6 & -2.04 & -2.33 & -3.65 & 1.29 & -3.69 & -2.21 & -14.76 & -1.47\tabularnewline
 & 256\_5 & -2.09 & -2.31 & -3.66 & 0.79 & -3.83 & -2.30 & -14.49 & -2.32\tabularnewline
 & 128\_7 & 2.44 & 1.67 & 0.79 & 8.73 & -3.54 & -2.43 & -13.04 & -0.71\tabularnewline
 & 128\_6 & 2.71 & 1.06 & 5.88 & 9.39 & -2.97 & -2.19 & -10.49 & -0.14\tabularnewline
 & 64\_7 & 2.05 & 0.76 & 1.44 & 10.35 & -2.48 & -1.93 & -9.30 & 1.02\tabularnewline
 & 64\_6 & 1.67 & 1.07 & -0.17 & 7.08 & -2.10 & -1.76 & -6.97 & 0.69\tabularnewline
\hline 
C\_RaceHorses & 512\_5 & -3.21 & -1.67 & -7.12 & -8.50 & -4.04 & -0.71 & -11.43 & -16.63\tabularnewline
 & 512\_4 & -2.58 & -1.39 & -5.92 & -6.38 & -3.68 & -0.88 & -10.12 & -14.03\tabularnewline
 & 256\_6 & -2.64 & -1.57 & -5.57 & -6.14 & -3.58 & -1.28 & -8.67 & -12.28\tabularnewline
 & 256\_5 & -2.52 & -1.55 & -5.49 & -5.36 & \textbf{-4.07} & \textbf{-1.53} & \textbf{-9.69} & \textbf{-13.70}\tabularnewline
 & 128\_7 & -0.58 & 0.61 & -2.82 & -5.51 & -3.81 & -1.57 & -8.91 & -12.19\tabularnewline
 & 128\_6 & -0.58 & 0.31 & -2.59 & -3.88 & -3.48 & -1.52 & -7.04 & -11.73\tabularnewline
 & 64\_7 & 0.39 & 0.38 & 0.45 & 0.42 & -2.18 & -1.39 & -2.25 & -6.86\tabularnewline
 & 64\_6 & 0.10 & 0.31 & -0.56 & -0.48 & -2.07 & -1.24 & -3.74 & -5.35\tabularnewline
\hline 
\end{tabular}
\end{table*}

\subsection{Results}


\subsubsection{Overview}

In Table \ref{tab:BD-Rate-and-BD-SNR-1}, the performance of the proposed
method is compared against that of the conventional VVC-based coding
method. The result of the pretrained filters without adaptation are
also shown for comparison. The results are reported in terms of Bj\o ntegaard
Delta-rate (BD-rate) criterion \citep{bjontegaard2001calculation}
for luminance (Y), chroma components (U and V), and a weighted average
over the three channels (YUV). For combined YUV BD-rate, in order
to compute YUV-SNR, a weighting of (6, 1, and 1) is used for Y, U,
and V channels, respectively. In case of VVC, the bitrate and PSNR
values reported by VTM are used to compute BD-rate. To report the
performance of the proposed method, the bitrate of finetuned network
is computed by adding the weight-update signaling overhead to the
bitstream size achieved from VTM, and PSNR is calculated between the
content processed by the finetuned neural network model and the uncompressed
source content. The pretrained network is considered part of the codec so
there is no extra overhead caused by the pretrained filter.

As can be seen from the table, apart from some cases with more complex architectures (i.e., using more blocks and convolutional filters), usually a pretrained filter
cannot provide any improvement on BD-rate. Although a general pretrained
filter does not incur any extra overhead, it cannot effectively improve
the BD-rate performance. After adaptation, all filters can improve
the BD-rate on all the video sequences. For high resolution group
B video sequence, the more complicated the network (with higher $N_{blocks}$ and
$N_{filters}$), the better improvement in BD-rate. This trend can
also be observed in the performance of pretrained filter. It reflects
that a complex filter can show better noise-removing ability than
a simpler one. The best performance is obtained with a postprocessing
filter with $N_{blocks}=5$ and $N_{filters}=512$ on BQTerrace dataset
which brings up to 9.7\%, 6.8\%, 17.1\% and 20.2\% gain in BD-rate
over combined YUV, Y, U, and V channels, respectively. The performance
also depends on the video sequence itself, varying from $4.4\%$ to
$9.7\%$. There are also significant differences between luminance
channel and chroma channel: the BD-rate gain of chroma channel can be
up to $20\%$, while that of luminance is up to $6.8\%.$ 

For lower resolution video, the effect of the extra finetuning overhead
is more significant so it is preferable to use architectures which introduce a smaller
overhead. Usually a smaller filter (for example $N_{filters}=128$)
gets a similar result as the larger one due to the smaller overhead.
In some situations (C\_BQMall and C\_BasketBallDrill), a complex pretrained
network can outperform the finetuning methodology in BD-rate as pretraining-only
approach does not introduce any overhead.

\begin{table*}
\caption{\label{tab:PSNR-and-the-1}PSNR and the bitrate of the first 49 frames
in two video sequences (B\_BQTerrace and C\_PartyScene). Anchor is
the VTM 7.0 result and Test is the result after filtering with either
128\_7 or 512\_5 network. The \% increase refers to the ratio between
the absolute increase with the anchor bitrate.}

B\_BQTerrace{

\begin{tabular}{|cc|cc|cc|cc|cc|}
\hline 
 & QP & \multicolumn{2}{c|}{22} & \multicolumn{2}{c|}{27} & \multicolumn{2}{c|}{32} & \multicolumn{2}{c|}{37}\tabularnewline
 & Network & 128\_7 & 512\_5 & 128\_7 & 512\_5 & 128\_7 & 512\_5 & 128\_7 & 512\_5\tabularnewline
\hline 
Bitrate & Anchor & 36469 & 36469 & 7230 & 7230 & 2444 & 2444 & 1102 & 1102\tabularnewline
(kps) & Test & 36502 & 36549 & 7262 & 7310 & 2477 & 2524 & 1135 & 1182\tabularnewline
 & Increase & 32.82 & 80.21 & 32.8 & 80.41 & 32.89 & 80.36 & 32.8 & 80.26\tabularnewline
 & \% Increase & 0.09 & 0.22 & 0.45 & 1.11 & 1.35 & 3.29 & 2.98 & 7.28\tabularnewline
\hline 
Average & Anchor & 39.15 & 39.15 & 36.96 & 36.96 & 35.42 & 35.42 & 33.68 & 33.68\tabularnewline
PSNR & Test & 39.19 & 39.24 & 37.04 & 37.1 & 35.56 & 35.62 & 33.86 & 33.93\tabularnewline
(dB) & Gain & 0.04 & 0.08 & 0.09 & 0.14 & 0.14 & 0.2 & 0.18 & 0.25\tabularnewline
\hline 
Y-PSNR & Anchor & 37.95 & 37.95 & 35.62 & 35.62 & 34.07 & 34.07 & 32.3 & 32.3\tabularnewline
(dB) & Test & 37.99 & 38.02 & 35.71 & 35.75 & 34.2 & 34.26 & 32.47 & 32.54\tabularnewline
 & Gain & 0.04 & 0.07 & 0.09 & 0.14 & 0.14 & 0.19 & 0.17 & 0.23\tabularnewline
\hline 
U-PSNR & Anchor & 42.69 & 42.69 & 41.6 & 41.6 & 40.19 & 40.19 & 38.64 & 38.64\tabularnewline
(dB) & Test & 42.7 & 42.82 & 41.61 & 41.8 & 40.31 & 40.5 & 38.89 & 39.12\tabularnewline
 & Gain & 0.01 & 0.13 & 0.02 & 0.2 & 0.12 & 0.31 & 0.25 & 0.48\tabularnewline
\hline 
V-PSNR & Anchor & 44.77 & 44.77 & 43.67 & 43.67 & 42.34 & 42.34 & 40.91 & 40.91\tabularnewline
(dB) & Test & 44.81 & 44.96 & 43.73 & 43.91 & 42.51 & 42.67 & 41.22 & 41.36\tabularnewline
 & Gain & 0.04 & 0.19 & 0.07 & 0.24 & 0.17 & 0.33 & 0.31 & 0.44\tabularnewline
\hline 
\end{tabular}

}

C\_PartyScene{

\begin{tabular}{|cc|cc|cc|cc|cc|}
\hline 
 & QP & \multicolumn{2}{c|}{22} & \multicolumn{2}{c|}{27} & \multicolumn{2}{c|}{32} & \multicolumn{2}{c|}{37}\tabularnewline
 & Network & 128\_7 & 512\_5 & 128\_7 & 512\_5 & 128\_7 & 512\_5 & 128\_7 & 512\_5\tabularnewline
\hline 
Bitrate & Anchor & 10220 & 10220 & 4987 & 4987 & 2554 & 2554 & 1279 & 1279\tabularnewline
(kps) & Test & 10253 & 10300 & 5020 & 5068 & 2587 & 2634 & 1312 & 1360\tabularnewline
 & Increase & 32.86 & 79.93 & 32.76 & 80.11 & 32.93 & 80.20 & 32.97 & 80.34\tabularnewline
 & \% Increase & 0.32 & 0.78 & 0.66 & 1.61 & 1.29 & 3.14 & 2.58 & 6.28\tabularnewline
\hline 
Average & Anchor & 39.46 & 39.46 & 36.03 & 36.03 & 33.04 & 33.04 & 30.23 & 30.23\tabularnewline
PSNR & Test & 39.55 & 39.60 & 36.20 & 36.26 & 33.24 & 33.30 & 30.42 & 30.47\tabularnewline
(dB) & Gain & 0.08 & 0.14 & 0.17 & 0.23 & 0.20 & 0.26 & 0.19 & 0.24\tabularnewline
\hline 
Y-PSNR & Anchor & 38.52 & 38.52 & 34.89 & 34.89 & 31.79 & 31.79 & 28.87 & 28.87\tabularnewline
(dB) & Test & 38.60 & 38.64 & 35.05 & 35.10 & 31.98 & 32.03 & 29.04 & 29.09\tabularnewline
 & Gain & 0.09 & 0.12 & 0.16 & 0.21 & 0.19 & 0.23 & 0.17 & 0.22\tabularnewline
\hline 
U-PSNR & Anchor & 42.04 & 42.04 & 39.79 & 39.79 & 37.72 & 37.72 & 36.10 & 36.10\tabularnewline
(dB) & Test & 42.25 & 42.39 & 40.17 & 40.32 & 38.23 & 38.39 & 36.65 & 36.79\tabularnewline
 & Gain & 0.21 & 0.34 & 0.39 & 0.53 & 0.52 & 0.68 & 0.55 & 0.69\tabularnewline
\hline 
V-PSNR & Anchor & 42.72 & 42.72 & 40.20 & 40.20 & 37.94 & 37.94 & 35.82 & 35.82\tabularnewline
(dB) & Test & 42.65 & 42.79 & 40.23 & 40.40 & 38.04 & 38.19 & 35.98 & 36.11\tabularnewline
 & Gain & -0.07 & 0.07 & 0.03 & 0.20 & 0.10 & 0.26 & 0.15 & 0.29\tabularnewline
\hline 
\end{tabular}}
\end{table*}

\begin{figure}
\includegraphics[width=0.9\columnwidth]{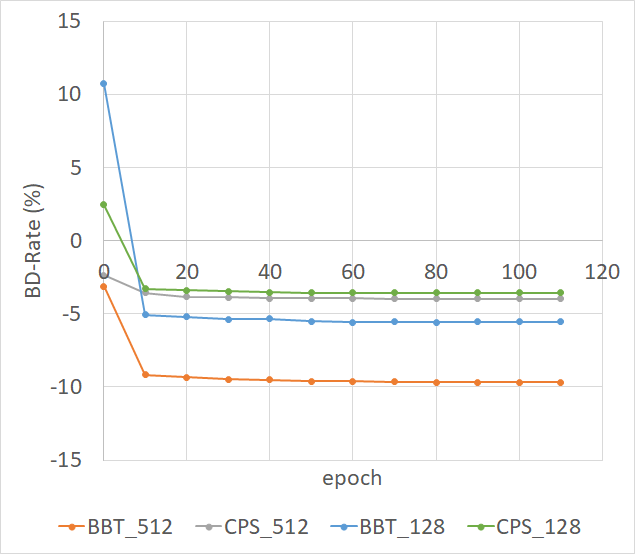}

\caption{\label{fig:BD-rate-epoch-plot-of}BD-rate-epoch plot of different
video sequences with the two neural networks ($N_{blocks}=7,N_{filters}=128$
and $N_{blocks}=5,N_{filters}=512$) on two different video sequences
C\_PartyScene (CPS) and B\_BQTerrace (BBT)}
\end{figure}

\subsubsection{Analysis}

A detailed analysis (Table \ref{tab:PSNR-and-the-1}) is done on the
first 49 frames of two video sequences B\_BQTerrace and C\_PartyScene
with two different neural networks ($N_{blocks}=7,N_{filters}=128$
and $N_{blocks}=5,N_{filters}=512$). It shows the bitrate and PSNR
values of video sequences for the anchor VTM result and our methodology
over the four tested rate/quality points. The weight-update overhead
percentage with respect to the VTM/Anchor bitrate is also shown in
Table \ref{tab:PSNR-and-the-1}. For bitrate, the results show that
our proposed method introduces higher weight-update overhead percentage in lower bitrates (i.e., higher QPs). However, for these cases, also the PSNR gains are higher, as there are more significant artifacts in the input data. Furthermore, the results indicate that the improvement in PSNR
decreases in higher bitrates where the task of removing compression
artifacts becomes more challenging for the neural network model, as artifacts are less significant. Although
only the result related to two sequences are presented in Table \ref{tab:PSNR-and-the-1},
a similar behavior is observed for the other tested video sequences.

The BD-rate gains in different epochs are shown in Figure \ref{fig:BD-rate-epoch-plot-of}. When the BD-rate keep decreasing in finetuning, a sufficient
finetuning convergence can be achieved in a short time. The figure shows
that a significant improvement can be found in the first 10 epochs.
It is an expected result as only a small proportion of weights was updated
in finetuning stage.

\begin{table*}
\caption{\label{tab:Time-complexity-of}Time complexity in second of different
processing stages of two different neural networks on different datasets.}

\begin{tabular}{|l|l|l|llll|}
\hline 
Network & Video sequence & Pretraining & VTM Encoding & Finetuning & VTM Decoding & Filtering\tabularnewline
\hline 
512\_5 & B\_BQTerrace\_37 & 1287435 & 5147 & 18980 & 3.5 & 63.3\tabularnewline
 & B\_BQTerrace\_22 &  & 26234 & 19152 & 8.6 & 63.1\tabularnewline
 \cline{4-7}
 & C\_PartyScene\_37 &  & 1608 & 3775 & 0.9 & 15.1\tabularnewline
 & C\_PartyScene\_22 &  & 5300 & 3789 & 1.8 & 15.1\tabularnewline
\hline 
128\_7 & B\_BQTerrace\_37 & 219711 & 5147 & 4397 & 3.5 & 16.4\tabularnewline
 & B\_BQTerrace\_22 &  & 26234 & 4331 & 8.6 & 16.0\tabularnewline
 \cline{4-7}
 & C\_PartyScene\_37 &  & 1608 & 1035 & 0.9 & 4.5\tabularnewline
 & C\_PartyScene\_22 &  & 5300 & 1056 & 1.8 & 5.0\tabularnewline
\hline 
\end{tabular}

\end{table*}

\subsubsection{Complexity}

To study the complexity of our methodology, the whole pipeline is
broken down into five steps. The time measurements are reported in Table \ref{tab:Time-complexity-of}. Pretraining time is the time required
to train the filter from scratch with the pretraining dataset (CLIC
dataset) on one single Nvidia Tesla V100 core. For fair comparison, both networks
are trained for 1000 epochs. The pretraining time is very long, however the
same network is used as the pretrained model for different test video
sequences.

The VTM encoding process is done on a single CPU of Intel Xeon Gold
6154 without parallel processing. The processing time depends on the
quality of video sequences, which includes the resolution and also
QP.

Finetuning is performed on a single Nvidia Tesla V100 core for 110 epochs
for each video sequence. The required time does not depend on the
QP of the video but only the number of processed patches, which grows
with the resolution. While the required time seems to be long
compared to VTM encoding time, the finetuning
can be stopped earlier without a significant drop of the adapting
performance. The required time for each epoch is nearly constant so
the finetuning time can be reduced to \textasciitilde$9\%${} of the
reported time if the finetuning is stopped at 10 epochs. This can provide extra
flexibility for different applications depending on the processing
environment.

\section{Conclusions}

In this paper, we presented a new methodology to apply an adaptive post-processing
filter for traditional video codecs. A pretrained neural network was
trained offline on a generic image dataset. At encoding stage, we
demonstrated the efficient adaptation of the neural network to the
target video content by finetuning only the bias terms of the CNN.
We evaluated the proposed idea on 7 different video sequences and
obtain 9.7\% in BD-rate gain on average compared to VVC VTM 7.0 video sequences.

\bibliographystyle{ACM-Reference-Format}
\bibliography{acmmm_ref.bib}

\end{document}